\documentclass[prd,twocolumn,nofootinbib,showpacs]{revtex4}
\usepackage{graphicx}
\usepackage{latexsym}

\begin{document}

\title{Constraint on Coupled Dark Energy Models from Observations}

\author{Jun-Qing Xia}
\email{xia@sissa.it}

\affiliation{Scuola Internazionale Superiore di Studi Avanzati, Via
Beirut 2-4, I-34014 Trieste, Italy}


\begin{abstract}

The coupled dark energy models, in which the quintessence scalar
field nontrivially couples to the cold dark matter, have been
proposed to explain the coincidence problem. In this paper we study
the perturbations of coupled dark energy models and the effects of
this interaction on the current observations. Here, we pay
particular attention to its imprint on the late-time Integrated
Sachs-Wolfe (ISW) effect. We perform a global analysis of the
constraints on this interaction from the current observational data.
Considering the typical exponential form as the interaction form, we
obtain that the strength of interaction between dark sectors is
constrained as $\beta<0.085$ at $95\%$ confidence level.
Furthermore, we find that future measurements with smaller error
bars could improve the constraint on the strength of coupling by a
factor two, when compared to the present constraints.

\end{abstract}


\maketitle


\section{Introduction}

Current cosmological observations, such as the cosmic microwave
background (CMB) measurements of temperature anisotropies and
polarization \cite{Komatsu:2008hk} and the redshift-distance
measurements of Type Ia Supernovae (SNIa) at $z<2$
\cite{Kowalski:2008ez}, have demonstrated that the Universe is now
undergoing an accelerated phase of expansion and that its total
energy budget is dominated by the dark energy component. The nature
of dark energy is one of the biggest unsolved problems in modern
physics and has been extensively investigated in recent years, both
under the theoretical and the observational point of view.

The simplest candidate of dark energy is the cosmological constant,
where the equation of state $w$ always remains $-1$. In the standard
$\Lambda$-Cold Dark Matter ($\Lambda$CDM) cosmology, cold dark
matter only interacts with other components by gravity, while dark
energy is simply a cosmological constant without any evolution and
perturbation. Although this concordance model can fit the current
observational data very well \cite{Komatsu:2008hk,Kowalski:2008ez},
the possibility of dynamical dark energy models cannot be ruled out
yet. Furthermore, this model suffers of the fine-tuning and
coincidence problems, i.e. why the Universe is dominated by dark
energy in late times \cite{CCproblem1,CCproblem2}.

In order to lift these severe tensions, many alternative dynamical
dark energy models, such as quintessence
\cite{Quint1,Quint2,Quint3,Quint4}, phantom \cite{Phantom},
K-essence \cite{kessence1,kessence2} and quintom \cite{Feng:2004ad},
have been proposed recently (for a review see Ref.\cite{copeland}).
For example, the quintessence model, a scalar field $\phi$ slowly
rolling down a potential energy $V(\phi)$, has the spatial
fluctuations and its equation of state $w_{\phi}$ can evolve with
the cosmic time. More importantly, being a dynamical component, the
quintessence is naturally expected to interact with the other
components, such as the cold dark matter
\cite{Copeland98,Amendola00} or massive neutrinos
\cite{gu03,fardon04}, in the field theory framework. If these
interactions really exist, it would open up the possibility of
detecting the dark energy non-gravitationally.

In the coupled dark energy models, the quintessence scalar field
could nontrivially couple to the cold dark matter component. The
presence of the interaction clearly modifies the cosmological
background evolutions. The evolution of cold dark matter energy
density $\rho^{}_{c}(a)$ will be different, when compared to that of
the minimally coupled models, and dependent on the quintessence
field $\phi\,$:
\begin{equation}
\rho^{}_{\rm c}(a)=\rho^{}_{\rm c0}a^{-3+\delta(\phi)}~,\label{cdm}
\end{equation}
where $a$ is the scale factor, $\rho^{}_{\rm c0}$ is the present
value of cold dark matter energy density, and $\delta(\phi)$ is the
modification due to the interactions and could be non-zero during
the evolution of Universe. In this case, at early times there will
be more (less) cold dark matter energy density, when
$\delta(\phi)<0$ ($>0$).

On the other hand, the interaction between dark sectors will also
affect the evolution of cosmological perturbations, which has been
widely investigated recently (see e.g.
Refs.\cite{hwangnoh02,amendola04,koivisto05,lee06,
olivares06,brookfield08,bean08,mainini08,valiviita08,he09a,chong09}).
Due to the different evolution of cold dark matter energy density,
the interaction between dark sectors could shift the
matter-radiation equality scale factor $a_{\rm eq}$, and affect the
locations and amplitudes of acoustic peaks of CMB temperature
anisotropies and the turnover scales of large scale structure (LSS)
matter power spectrum consequently. In addition, the interaction
will also affect the late ISW effect \cite{isw} at large scales
which is produced by the CMB photons passing through the
time-evolving gravitational potential well, when dark energy or
curvature becomes important at later times
\cite{brookfield08,he09b}.

Therefore, with the accumulation of observational data and the
improvements of the data quality, it is of great interest to
investigate the non-minimally coupled dark energy models from the
current observational data. In this paper we study the perturbations
of coupled dark energy models and present the constraints on the
interactions from the observational data in detail. The structure of
the paper is as follows: in Sec.II and Sec.III we show the basic
equations of background evolution and linear perturbations of the
coupled dark energy model, respectively. In Sec.IV we present the
current and future observational datasets we used. Sec.V contains
our main global fitting results from the current and future
observations, while Sec.VI is dedicated to the summary.


\section{Background Evolution}

In our analysis we assume a flat, homogeneous,
Friedmann-Robertson-Walker universe with metric
\begin{equation}
ds^2=a^2(\eta)\left(-d\eta^2+dx_idx^i\right)~,
\end{equation}
where $\eta$ is the conformal time. The Friedmann equation is:
\begin{equation}
\mathcal{H}^2\equiv\left(\frac{\dot{a}}{a}\right)^2=\frac{8\pi
G}{3}a^2\rho^{}_{\rm tot}~,\label{friedeq}
\end{equation}
where $\rho^{}_{\rm tot}$ is the total energy density of Universe
and the dot refers to the derivative with respect to the conformal
time $\eta$.

From the Lagrangian of the quintessence scalar field:
\begin{equation}
\mathcal{L}_{\phi}=\frac{1}{2}\partial_{\mu}\phi\,\partial^{\mu}\phi-V(\phi)~,
\end{equation}
we can obtain the energy density and pressure
\begin{equation}
\rho^{}_{\phi}=\frac{\dot{\phi}^2}{2a^2}+V(\phi)~,~~~p^{}_{\phi}=\frac{\dot{\phi}^2}{2a^2}-V(\phi)~.\label{rhop}
\end{equation}
And the Friedmann equation Eq.(\ref{friedeq}) becomes:
\begin{equation}
\mathcal{H}^2=\frac{8\pi G}{3}a^2\left(\rho^{}_{\gamma}+\rho^{}_{\rm
b}+\rho^{}_{\rm c}+\frac{\dot{\phi}^2}{2a^2}+V(\phi)\right)~,
\end{equation}
where $\rho^{}_{\gamma}$, $\rho^{}_{\rm b}$ and $\rho^{}_{\rm c}$
are the energy densities of radiation, baryons and cold dark matter,
respectively. Consequently, we define the energy density parameters
$\Omega_{\,\rm i}\equiv\rho_{\rm i}/\rho_{\rm tot}$, where
$\rho_{\rm i}$ is the energy density of each component. In the
following calculations we will set the reduced Planck mass $M_{\rm
pl}=1/\sqrt{8\pi G}\equiv1$.

Here, we consider the typical exponential form as the quintessence
potential, namely
\begin{equation}
V(\phi)=V_{0}e^{-\lambda\phi}~.\label{potential}
\end{equation}
We also take the exponential form:
\begin{equation}
\rho^{}_{\rm c}(\phi)=\rho^{\ast}_{\rm
c}e^{\beta\phi}~,\label{coupling}
\end{equation}
as the interaction form between quintessence and cold dark matter,
where $\rho^{\ast}_{\rm c}$ is the bare energy density of cold dark
matter and $\beta$ is the strength of interaction. There are several
stringent constrains on this strength of interaction, namely,
$|\beta|\lesssim\mathcal{O}(0.1)$ at $95\%$ confidence level, from
different observations \cite{bean08,amendola1,amendola2}. Using
these potential forms, the initial choice of $\phi_{\,\rm i}$ will
not affect the evolution of Universe significantly. And this coupled
model could be reduced to the standard $\Lambda$CDM model by
choosing $\lambda=\beta=0$ straightforwardly. The effective
potential of this coupled system is given by: $V_{\rm
eff}(\phi)=V(\phi)+\rho^{}_{\rm c}(\phi)$. If we restrict to the
case $\lambda>0$, this effective potential will have a minimal value
at $\phi_{\rm min}=\ln{(\lambda V_0/\beta\rho^{}_{\rm c})}/\lambda$
for the strength $\beta>0$. Therefore, in our following calculations
we will restrict the forms to the case $\lambda>0$ and $\beta>0$ to
lead to the acceleration expansion of Universe at late times.

When including the interactions, the conservation of energy momentum
for each component becomes \cite{brookfield08}:
\begin{equation}
T^{\mu}_{\nu;\mu}=\beta\phi_{,\nu}T^{\alpha}_{\alpha}~.\label{tmunu}
\end{equation}
In our analysis, we assume that the baryons and radiation are not
coupled with the quintessence scalar field. Therefore, when
considering the interaction between cold dark matter and
quintessence, the energy conservation equations of CDM and
quintessence will be violated, while those of baryons and radiation
are held:
\begin{eqnarray}
\dot{\rho}^{}_{\gamma}+4\mathcal{H}{\rho}^{}_{\gamma}&=&0~,\\
\dot{\rho}^{}_{\rm b}+3\mathcal{H}{\rho}^{}_{\rm b}&=&0~,\\
\dot{\rho}^{}_{\rm c}+3\mathcal{H}{\rho}^{}_{\rm c}&=&Q_{\rm c}=\beta\dot{\phi}\rho^{}_{\rm c}~,\\
\dot{\rho}^{}_{\phi}+3\mathcal{H}{\rho}^{}_{\phi}(1+w^{}_{\phi})&=&Q_{\phi}=-\beta\dot{\phi}\rho^{}_{\rm
c}~,\label{decoveq}
\end{eqnarray}
where the energy exchange $Q_{\rm c}$ is the function of $\phi$.
When $Q_{\rm c}<0$ ($>0$), the energy of cold dark matter (dark
energy) transfers to
dark energy (cold dark matter). 
Because we describe the dark energy component using a quintessence
scalar field, the energy conservation equation Eq.(\ref{decoveq})
will be describe as the Klein-Gordon equation:
\begin{equation}
\ddot{\phi}+2\mathcal{H}\dot{\phi}+a^2V'(\phi)=-a^2\beta\rho^{}_{\rm
c}~,
\end{equation}
where the prime denotes the derivative with respect to the
quintessence scalar field $\phi\,$: $V'(\phi)\equiv\partial
V(\phi)/\partial\phi\,$.

We can also define the effective equation of state of CDM and
quintessence field to describe the equivalent uncoupled model in the
background:
\begin{eqnarray}
\dot{\rho}^{}_{\rm c}+3\mathcal{H}{\rho}^{}_{\rm
c}(1+w^{\rm eff}_{\rm c})&=&0~,\\
\dot{\rho}^{}_{\phi}+3\mathcal{H}{\rho}^{}_{\phi}(1+w^{\rm
eff}_{\phi})&=&0~,
\end{eqnarray}
where the effective equation of state are given by:
\begin{equation}
w^{\rm eff}_{\rm c}=-\frac{\beta\dot{\phi}}{3\mathcal{H}}~,~~~
w^{\rm
eff}_{\phi}=w_{\phi}+\frac{\beta\dot{\phi}}{3\mathcal{H}}\frac{\rho_{\rm
c}}{\rho_{\phi}}~.\label{effeos}
\end{equation}
We can find that in this case the effective equation of state of
cold dark matter $w^{\rm eff}_{\rm c}$ will be non-vanishing and
evolves with the cosmic time.

\begin{figure}[t]
\begin{center}
\includegraphics[scale=0.45]{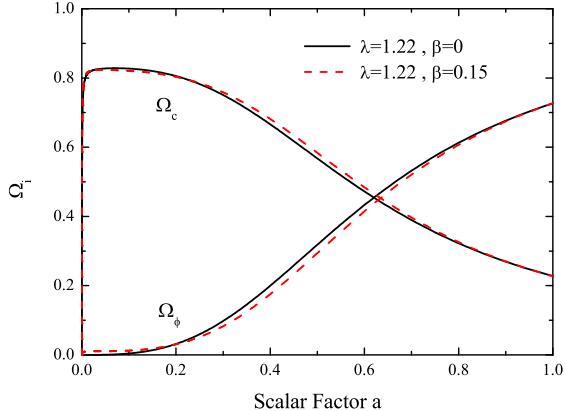}
\caption{The cosmological evolutions of the cold dark matter and
quintessence energy density parameters, $\Omega_{\,\rm i}$, for two
models: $\lambda=1.22$, $\beta=0$ (black solid lines) and
$\lambda=1.22$, $\beta=0.15$ (red dashed lines).\label{fig1}}
\end{center}
\end{figure}

Based on these equations, we can study the background evolutions of
each component in the non-minimal coupling system. In Fig.\ref{fig1}
we illustrate the evolutions of the cold dark matter and
quintessence energy density parameters as a function of redshift for
two different models: the uncoupled model $\lambda=1.22$, $\beta=0$
(black solid lines) and the coupled system $\lambda=1.22$,
$\beta=0.15$ (red dashed lines). For other cosmological parameters,
we fix them to the best fit values of WMAP5 data
\cite{Komatsu:2008hk}: $\Omega_{\rm b}h^2=0.02267$, $\Omega_{\rm
c}h^2=0.1131$, and $h=0.705$. Here, we normalize energy density
parameters to their present values.

Clearly, the interaction between the quintessence scalar field and
the cold dark matter significantly modifies the evolutions of their
energy density. Because of the presence of the coupling, during the
evolution, $\beta\dot{\phi}<0$ and the energy density of cold dark
matter evolves faster than $a^{-3}$. Therefore, if we normalize
energy density parameters to their present values, the cold dark
matter energy density will be larger than that of the uncoupled
model. The energy of cold dark matter transfers to dark energy.


\section{Linear Perturbation}

In this section we will consider the evolution of cosmological
perturbations in the non-minimally coupled quintessence model.
Starting from the metric in the synchronous gauge
\cite{pertks84,pertrev,ma95}, the line element is given by:
\begin{equation}
ds^2=-a^2(\eta)\left(-d\eta^2+(\delta_{ij}+h_{ij})dx^idx^j\right)~,
\end{equation}
where $h_{ij}$ is the metric perturbation in this synchronous
coordinate system. Here, we restrict only for the scalar mode of the
metric perturbations.

We write the inhomogeneous energy density of cold dark matter and
the quintessence scalar field as:
\begin{eqnarray}
\rho^{}_{\rm c}(\eta,\vec{x})&=&\rho^{}_{\rm
c}(\eta)+\delta\rho^{}_{\rm
c}(\eta,\vec{x})\nonumber\\
&=&\rho^{}_{\rm c}(\eta)(1+\delta^{}_{\rm
c}(\eta,\vec{x}))~,\\
\phi(\eta,\vec{x})&=&\phi(\eta)+\delta\phi(\eta,\vec{x})~,
\end{eqnarray}
where $\rho^{}_{\rm c}(\eta)$ and $\phi(\eta)$ are the background
parts, while $\delta\rho^{}_{\rm c}$ and $\delta\phi$ are the
perturbations. Using the perturbed part of the energy momentum
conservation equation Eq.(\ref{tmunu}) for the coupled system, we
could calculate the evolution equations for the perturbation of cold
dark matter:
\begin{eqnarray}
\dot{\delta}_{\rm c}&=&3(\mathcal{H}+\beta\dot{\phi})\left(w_{\rm
c}-\frac{\delta p_{\rm c}}{\delta\rho_{\rm c}}\right)\delta_{\rm
c}-(1+w_{\rm c})\left(\theta_{\rm
c}+\frac{\dot{h}}{2}\right)\nonumber\\
&&+\beta(1-3w_{\rm
c})\delta\dot{\phi}+\beta'\dot{\phi}\,\delta\phi(1-3w_{\rm c})~,\\
\dot{\theta}_{\rm c}&=&-\mathcal{H}(1-3w_{\rm c})\theta_{\rm
c}-\frac{\dot{w_{\rm c}}}{1+w_{\rm c}}\theta_{\rm c}+\frac{\delta
p_{\rm c}/\delta\rho_{\rm c}}{1+w_{\rm c}}k^2\delta_{\rm
c}\nonumber\\
&&+\beta\frac{1-3w_{\rm c}}{1+w_{\rm
c}}k^2\delta\phi-\beta(1-3w_{\rm c})\dot{\phi}\theta_{\rm
c}-k^2\sigma_{\rm c}~,
\end{eqnarray}
where $\delta p_{\rm c}$ is the pressure perturbation of cold dark
matter, $h$ is the metric perturbation, $\theta^{}_{\rm
c}=ik^{}_jv^j_{\rm c}$ is the gradient of the velocity field, and
$\sigma_{\rm c }$ is the shear stress of cold dark matter. In our
coupling form Eq.(\ref{coupling}), $\beta$ is a constant, so we have
$\beta'\equiv\partial\beta/\partial\phi=0$. Finally we obtain the
perturbation equations of cold dark matter:
\begin{eqnarray}
\dot{\delta}_{\rm c}&=&-\theta_{\rm
c}-\dot{h}/2+\beta\delta\dot{\phi}~,\\
\dot{\theta}_{\rm c}&=&-\mathcal{H}\theta_{\rm
c}-\beta\dot{\phi}\theta_{\rm c}+k^2\beta\delta{\phi}~.
\end{eqnarray}
We note that in the presence of interaction, the gradient of the
velocity $\theta_{\rm c}$ will evolve to be nonzero, even if its
initial value is zero; whereas $\theta_{\rm c}$ remains zero at all
times in the minimally coupled model. Therefore, in our calculations
we will evolve the perturbation equations in an arbitrary
synchronous gauge, instead of the cold dark matter rest frame
\cite{bean08}.

Meanwhile, the perturbed Klein-Gordon equation is given by:
\begin{equation}
\delta\ddot{\phi}+2\mathcal{H}\delta\dot{\phi}+k^2\delta\phi+
a^2V''\delta\phi+\dot{h}\dot{\phi}/2=-a^2\beta\rho_{\rm
c}\delta_{\rm c}~.
\end{equation}
We find that the perturbations of this non-minimally coupled system
are stable. This is different from some coupling models in which
dark energy component is modelled as a fluid with constant equation
of state parameter $w$. In those models, when $w$ is close to the
cosmological constant boundary, the coupling term $Q\sim\rho_{\rm
c}$ will lead to an instability
\cite{valiviita08,he09a,chong09,Gavela09}.

\begin{figure*}[htbp]
\begin{center}
\includegraphics[scale=0.43]{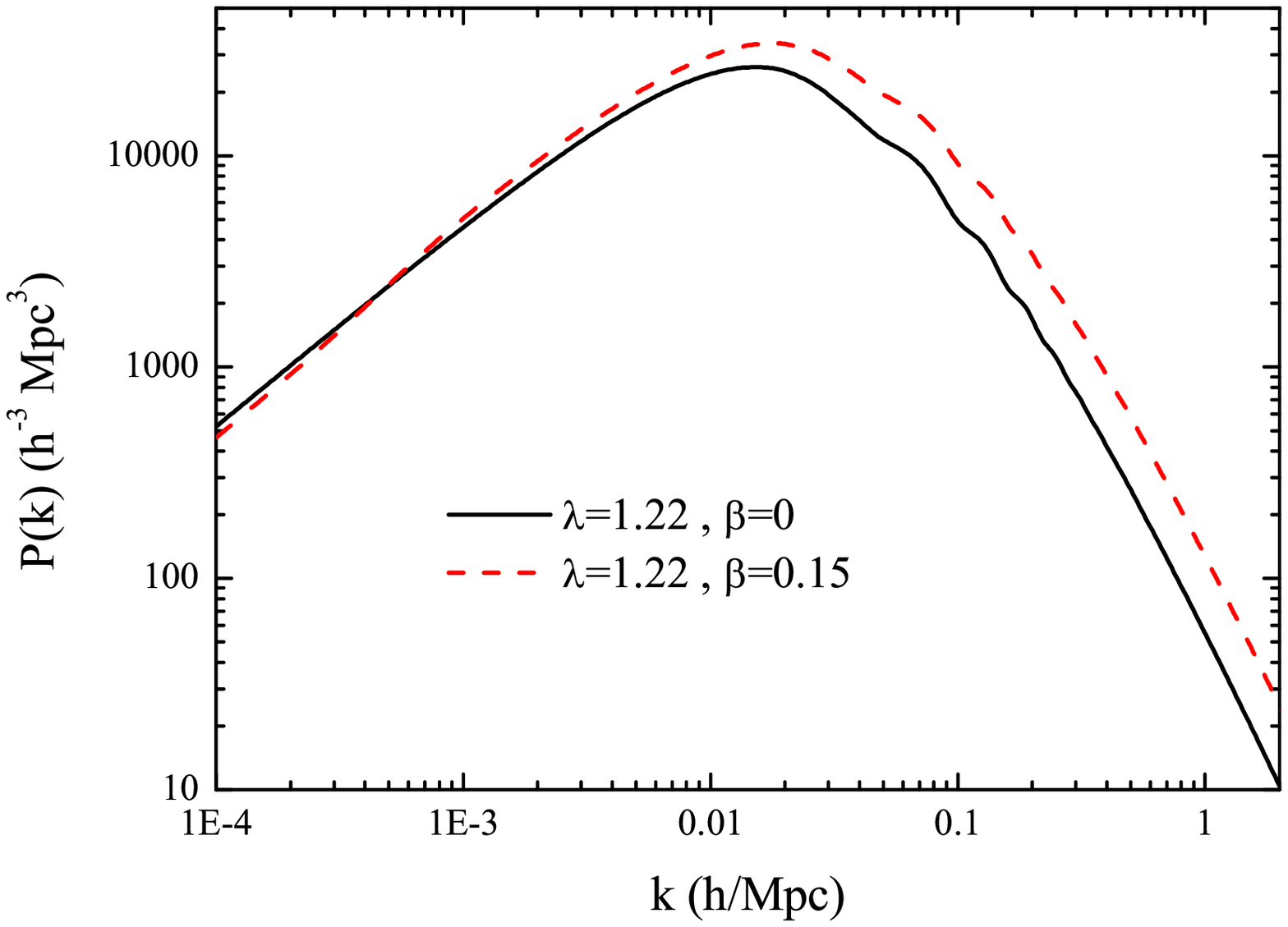}
\includegraphics[scale=0.43]{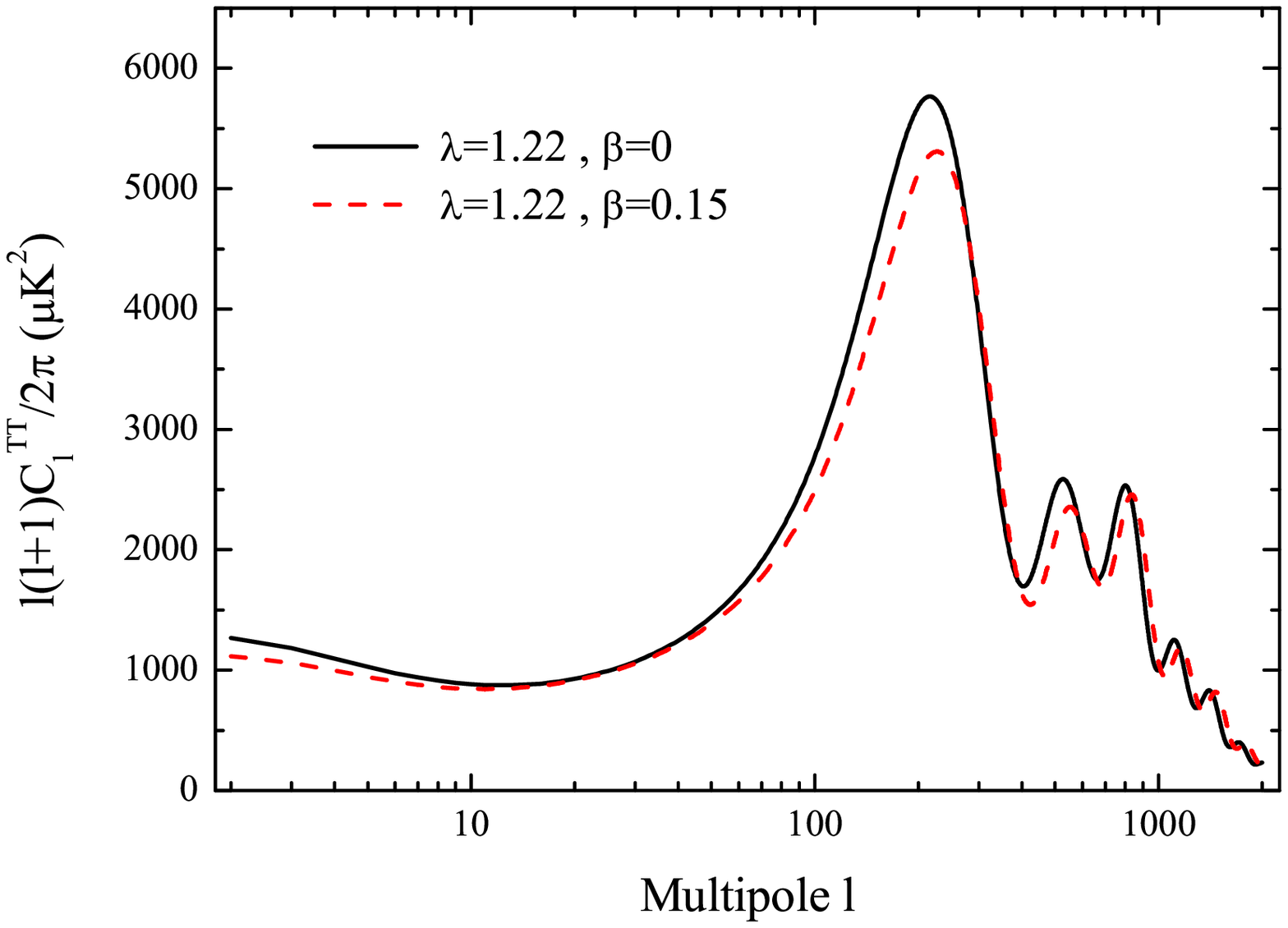}
\caption{The CMB temperature anisotropies and LSS matter power
spectrum for two models: $\lambda=1.22$, $\beta=0$ (black solid
lines) and $\lambda=1.22$, $\beta=0.15$ (red dashed
lines).\label{fig2}}
\end{center}
\end{figure*}

In this paper we still choose the adiabatic initial condition, which
implies that the entropy perturbation between cold dark matter and
quintessence:
\begin{equation}
\mathcal{S}=\frac{\delta\rho^{}_{\rm c}}{\rho^{}_{\rm
c}}-\frac{\delta\rho^{}_{\phi}}{\rho^{}_{\phi}}=0~,
\end{equation}
and the quintessence intrinsic entropy perturbation:
\begin{equation}
\mathcal{I}=\frac{\delta\rho^{}_{\phi}}{\rho^{}_{\phi}}-\frac{\delta
p^{}_{\phi}}{p^{}_{\phi}}=0~,
\end{equation}
vanish at early times. However, during the evolution of Universe,
the iso-curvature perturbations will be produced and the adiabatic
condition is generally no longer conserved, even for large-scale
modes, due to the presence of non-minimal coupling
\cite{chong09,valiviita08}.

Following the above equations, we modify the {\tt CAMB} code
\cite{camb} to calculate the CMB temperature power spectrum $C^{\rm
TT}_{\ell}$ and LSS matter power spectrum $P(k)$ in this non-minimal
coupling scenario. Besides the background parameters, we also set
$\tau=0.084$, $n_{\rm s}=0.96$ and $A_{\rm s}=2.15\times10^{-9}$ at
$k=0.05\,{\rm Mpc}^{-1}$, which are the best fit values obtained
from the WMAP5 data \cite{Komatsu:2008hk}.

In Fig.\ref{fig2} we plot the CMB temperature power spectrum and LSS
matter power spectrum for two different models. In this
non-minimally coupled system the matter power spectrum is almost
unchanged and the growth of density perturbations are not affected
significantly on the large scales \cite{lee06}. However, on the
small scales the situation will be different. As we mentioned
before, at early times there is higher cold dark matter energy
density in the coupled model with $\beta>0$. In this case, the epoch
of equality $a_{\rm eq}$ occurs further from the recombination. An
earlier $a_{\rm eq}$ means only the very small scale modes enter the
horizon and decay during the radiation dominated epoch. Therefore,
the amplitude of small scale matter power spectrum is enhanced and
the turnover occurs on the smaller scales as shown in the left panel
of Fig.\ref{fig2}. Consequently, the expected value of $\sigma_8$
will become larger than that of the minimally coupled case.

In the CMB temperature power spectrum, the most obvious effect is
that the amplitude on the small scales has been decreased
significantly due to the presence of interaction between dark
sectors. In the non-minimally coupled model, the epoch of equality
$a_{\rm eq}$ occurs further from the recombination, so that at
recombination the inhomogeneities induced from the radiation becomes
small. The small-scale anisotropies decrease when the coupling
strength is increasing. In addition, the locations of the acoustic
peaks in the CMB temperature anisotropies will also slightly shifted
to the smaller scales, due to the presence of interaction.

More interestingly, the non-minimal coupling $\beta>0$ also
suppresses the anisotropies on the large scales ($\ell < 20$) which
arise mainly from the late-time ISW effect. When $\beta$ is
positive, the energy transfers from cold dark matter to dark energy
and at early times cold dark matter density becomes larger.
Consequently, the dark matter-dark energy equality scale factor
$a'_{\rm eq}$ will be larger than that of the un-coupled case.
Because the gravitational potential $\Phi$ stays constant in the
matter dominated era, $\dot{\Phi}$ and additional CMB anisotropies
at large scales will be suppressed in the presence of non-minimal
coupling. The CMB temperature anisotropies on very large scales
become small, as shown in Fig.\ref{fig2}. Therefore, observing the
late time ISW effect could be a helpful way of studying the
nontrivially coupling between dark sectors.

\begin{figure}[t]
\begin{center}
\includegraphics[scale=0.45]{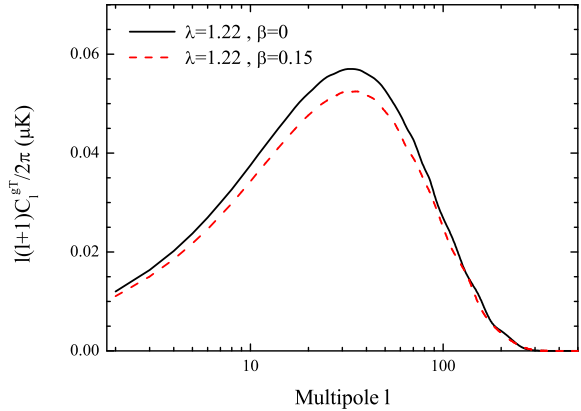}
\caption{The cross-correlation signal $C_{\ell}^{gT}$ for two
models: $\lambda=1.22$, $\beta=0$ (black solid lines) and
$\lambda=1.22$, $\beta=0.15$ (red dashed lines).\label{fig3}}
\end{center}
\end{figure}

However, the most significant ISW effect contributes to the CMB
anisotropies on large scales that are strongly affected by the
cosmic variance. Fortunately, this problem can be solved by the
cross-correlation between ISW temperature fluctuation and the
density of astrophysical objects like galaxies or quasars. (For the
details, we refer the reader to
Refs.\cite{peiris00,cooray02,afshordi04}.) In Fig.\ref{fig3} we plot
the cross-correlation power spectra $C_{\ell}^{gT}$ for two
different models using the public package {\tt
CAMB${}_{-}$sources}\footnote{Available at
http://camb.info/sources/.}. One can find that the power spectrum
$C_{\ell}^{gT}$ of the non-minimal coupling case will also be
suppressed. The non-minimal coupling leaves an interesting imprint
on the cross correlation power spectrum. We expect that the
observational ISW data could improve the constraints on the
interaction.

\section{Observational Data}
\subsection{Current Datasets}
In our calculations, we will rely here on the following cosmological
probes: i) CMB anisotropies and polarization; ii) baryonic acoustic
oscillations in the galaxy power spectra; iii) SNIa distance moduli;
and iv) the angular auto-correlation (ACF) and cross-correlation
(CCF) functions data obtained from the quasar map and CMB map.

In the computation of CMB power spectra we have included the WMAP
five-year (WMAP5) temperature and polarization power spectra with
the routines for computing the likelihood supplied by the WMAP team
\cite{Komatsu:2008hk,hinshaw09,gold09,WMAP5:Other1,WMAP5:Other2,WMAP5:Other3}.
Since the non-minimal coupling also leaves some signatures on the
small scales CMB power spectrum, in our calculations we also include
some small-scale CMB measurements to improve the constraints, such
as BOOMERanG \cite{boomerang}, CBI \cite{cbi} and ACBAR
\cite{acbar}.

BAOs (Baryonic Acoustic Oscillations) have been detected in the
current galaxy redshift survey data from the SDSS and the Two-degree
Field Galaxy Redshift Survey (2dFGRS)
\cite{Eisenstein:2005su,Cole:2005sx,Huetsi:2005tp,BAO}. The BAO can
directly measure not only the angular diameter distance, $D_A(z)$,
but also the expansion rate of the universe, $H(z)$, which is
powerful for studying dark energy \cite{Albrecht:2006um}. Since
current BAO data are not accurate enough for extracting the
information of $D_A(z)$ and $H(z)$ separately \cite{Okumura:2007br},
one can only determine an effective distance
\cite{Eisenstein:2005su}:
\begin{equation}
D_v(z)\equiv\left[(1+z)^2D_A^2(z)\frac{cz}{H(z)}\right]^{1/3}~.
\end{equation}
In this paper we use the Gaussian priors on the distance ratios
$r_s(z_d)/D_v(z)$:
\begin{eqnarray}
r_s(z_d)/D_v(z=0.20)&=&0.1980\pm0.0058~,\nonumber\\
r_s(z_d)/D_v(z=0.35)&=&0.1094\pm0.0033~,
\end{eqnarray}
with a correlation coefficient of $0.39$, extracted from the SDSS
and 2dFGRS surveys \cite{BAO}, where $r_s$ is the comoving sound
horizon size and $z_d$ is drag epoch at which baryons were released
from photons given by Ref.\cite{Eisenstein:1997ik}.

The SNIa data provide the luminosity distance as a function of
redshift which is also a very powerful measurement of cosmological
evolution. The supernovae data we use in this paper are the recently
released Union compilation (307 samples) from the Supernova
Cosmology project \cite{Kowalski:2008ez}, which include the recent
samples of SNIa from the (Supernovae Legacy Survey) SNLS and ESSENCE
survey, as well as some older data sets, and span the redshift range
$0\lesssim{z}\lesssim1.55$. In the calculation of the likelihood
from SNIa we have marginalized over the nuisance parameter as done
in Ref.\cite{SNMethod}.

For the observational ISW data, we use the ACF and CCF data obtained
from the quasar map of Sloan Digital Sky Survey Data Release Six
\cite{richards09} and the WMAP5 Internal Linear Combination (ILC)
map \cite{hinshaw09}, which has been used in Ref.\cite{xiaetal09}.
(Also, Refs.\cite{giannonho,hoetal} provided the observational ISW
data using different CMB and LSS surveys.) This ISW data set gives
the information at very low redshift ($z\sim1$) which is much lower
than the recombination ($z\sim1100$). Thus, we do not consider the
possible covariance matrix between this ISW data and the WMAP power
spectra $C_{\ell}$.

Furthermore, we make use of the Hubble Space Telescope (HST)
measurement of the Hubble parameter $H_{0}\equiv
100\,h$~km~s$^{-1}$~Mpc$^{-1}$ by a Gaussian likelihood function
centered around $h=0.72$ and with a standard deviation $\sigma=0.08$
\cite{HST}.

\subsection{Future Datasets}

In order to forecast future measurements we will use the same
observables as before without BAO.

For the simulation with Planck \cite{Planck}, we follow the method
given in Ref.\cite{xiaplanck} and mock the CMB temperature (TT) and
polarization (EE) power spectra and temperature-polarization
cross-correlation (TE) with the isotropic noise by assuming a given
fiducial cosmological model. In Table I, we list the assumed
experimental specifications of the future (mock) Planck measurement.
Here we neglect foregrounds and the tensor information. The
effective $\chi^2$ is:
\begin{eqnarray}
\chi_{\rm eff}^2 &\equiv& -2 \log \mathcal{L}\nonumber\\
&=&\sum_{\ell} (2{\ell}+1) \left \{ \log \left (
\frac{C_{\ell}^{TT}C_{\ell}^{EE}-(C_{\ell}^{TE})^2}{\tilde
C_{\ell}^{TT}\tilde
C_{\ell}^{EE}-(\tilde C_{\ell}^{TE})^2} \right )\right. \\
&&\left.+ \frac{\tilde
C_{\ell}^{TT}C_{\ell}^{EE}+C_{\ell}^{TT}\tilde C_{\ell}^{EE}-2\tilde
C_{\ell}^{TE}C_{\ell}^{TE}}{C_{\ell}^{TT}C_{\ell}^{EE}-(C_{\ell}^{TE})^2}
-2 \right \}~\nonumber,
\end{eqnarray}
where $C_{\ell}^{XY}$ denote theoretical power spectra and $\tilde
C_{\ell}^{XY}$ denote the power spectra from the simulated data. The
gaussian likelihood function has been normalized with respect to the
maximum likelihood, where $C_{\ell}^{XY}=\tilde C_{\ell}^{XY}$
\cite{Easther:2004vq,Perotto:2006rj}.

\begin{table}
TABLE I. Assumed experimental specifications for the mock
Planck-like measurements. The noise parameters $\Delta_T$ and
$\Delta_P$ are given in units of $\mu$K-arcmin.
\begin{center}
\begin{tabular}{cccccc}
\hline \hline

$f_{\rm sky}$~ & ~${\ell}_{\rm max}$~ & (GHz) &
~$\theta_{\rm fwhm}$~ & ~$\Delta_T$~~ & ~~$\Delta_P$~ \\

\hline

 0.65 & 2500 & 100 & 9.5' & 6.8 & 10.9 \\
      &      & 143 & 7.1' & 6.0 & 11.4 \\
      &      & 217 & 5.0' & 13.1 & 26.7 \\

\hline \hline
\end{tabular}
\end{center}
\end{table}

The proposed satellite SNAP (Supernova / Acceleration Probe) will be
a space based telescope with a one square degree field of view that
will survey the whole sky \cite{SNAP}. It aims at increasing the
discovery rate of SNIa to about $2000$ per year in the redshift
range $0.2<z<1.7$. In this paper we simulate about $2000$ SNIa
according to the forecast distribution of the SNAP
\cite{Kim:2003mq}. For the error, we follow the
Ref.\cite{Kim:2003mq} which takes the magnitude dispersion to be
$0.15$ and the systematic error $\sigma_{\rm sys}=0.02\times z/1.7$.
The whole error for each data is given by $\sigma_{\rm
mag}(z_i)=\sqrt{\sigma^2_{\rm sys}(z_i)+0.15^2/{n_i}}~$, where $n_i$
is the number of supernovae of the $i'$th redshift bin. Here, we do
not consider the possible covariance matrix among these redshift
bins.

For the future ISW ACF and CCF data, we simulate the mock dataset
from the best fit values of current data combination
CMB+BAO+SNIa+ISW. We reduce the error bars of these data points by a
factor three. We also use directly the covariance matrix taken from
present data and divide it by a factor nine. This improvement could
be achievable by next generation of large-scale surveys such as the
Large Synoptic Survey Telescope (LSST, \cite{Tyson02}), the
Panoramic Survey Telescope and Rapid Response System (Pan-STARRS,
\cite{Kaiser02}) and the Dark Energy Survey (DES; The Dark Energy
Survey Collaboration 2005). These surveys are likely to allow for an
order of magnitude improvement in the number of quasars, which
account for the factor three considered here if we assume that the
error bars scale as $\sqrt{N_{\rm quasar}}$. Also, the coverage of
the sky fraction is at present of the order of $20\%$ by SDSS DR6,
so an all-sky survey will already give a factor two improvement.
Another possible improvement in this direction is the use of type-2
quasars instead of the type-1 used here that could further decrease
the error bars of the sample (see the discussion in
Ref.\cite{richards09}).


\section{Numerical Results}

In our analysis, we perform a global fitting using the {\tt CosmoMC}
package \cite{cosmomc} a Monte Carlo Markov Chain (MCMC) code, which
has been modified to calculate the theoretical ACF and CCF. We
assume purely adiabatic initial conditions and a flat universe, with
no tensor contribution. Besides the parameters $\lambda$ and $\beta$
in the potential and coupling forms, we also vary the following
cosmological parameters with top-hat priors: the dark matter energy
density $\Omega_{\rm c} h^2 \in [0.01,0.99]$, the baryon energy
density $\Omega_{\rm b} h^2 \in [0.005,0.1]$, the primordial
spectral index $n_{\rm s} \in [0.5,1.5]$, the primordial amplitude
$\log[10^{10} A_{\rm s}] \in [2.7,4.0]$ and the angular diameter of
the sound horizon at last scattering $\theta \in [0.5,10]$. For the
pivot scale we set $k_{\rm s0}=0.05\,$Mpc$^{-1}$. When CMB data are
included, we also vary the optical depth to reionization $\tau \in
[0.01,0.8]$. We do not consider any massive neutrino contribution.
From the parameters above the MCMC code derives the reduced Hubble
parameter $H_0$, the present matter fraction $\Omega_{{\rm m}0}$,
and $\sigma_8$, so, these parameters have non-flat priors and the
corresponding bounds must be interpreted with some care. There are
three more parameters related to the ACF and CCF data: the constant
bias $b$, the efficiency of quasar catalog $a$ and the ISW amplitude
$A_{\rm amp}$ which is defined by: $\bar{C}^{\rm qT}(\theta)=A_{\rm
amp}C^{\rm qT}(\theta)$, where $\bar{C}^{\rm qT}$ and $C^{\rm qT}$
are the observed and theoretical CCF. In addition, {\tt CosmoMC}
imposes a weak prior on the Hubble parameter: $h \in [0.4,1.0]$.

\begin{figure}[t]
\begin{center}
\includegraphics[scale=0.45]{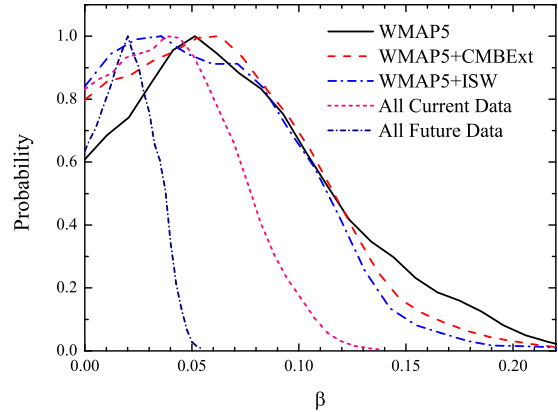}
\caption{One dimensional posterior distributions of the strength of
interaction $\beta$ from various data combinations: WMAP5 alone
(black solid line), WMAP5+CMBExt (red dashed line), WMAP5+ISW (blue
dash-dot line), all current data sets (magenta short dashed line),
and all future data sets (navy blue short dash-dot
line).\label{fig4}}
\end{center}
\end{figure}

In Fig.\ref{fig4} we show the one-dimensional posterior
distributions of $\beta$ from various data combinations. Firstly, we
present the constraint from the WMAP5 data alone (black solid line).
For the strength of interaction, the constraint is still weak,
namely the $95\%$ upper limit is $\beta<0.165$, which is consistent
with other results in the literature
\cite{bean08,amendola1,amendola2}. As we mentioned before, the
non-minimal coupling shifts the acoustic peaks of CMB temperature
anisotropies on the small scales. The small-scale CMB measurements
(CMBExt) should improve the constraints on the coupling strength.
Therefore, we plot the constraint on $\beta$ from WMAP5+CMBExt data
combination in Fig.\ref{fig4} (red dashed line). When adding the
small-scale CMB data, we can find that the constraint on $\beta$
improves slightly: $\beta<0.135$ at $95\%$ confidence level. The
small-scale CMB observations indeed improve the constraint.

We also include the observational ISW data \cite{xiaetal09} into our
calculations. In Fig.\ref{fig4} we show the one dimensional
distribution of $\beta$ from WMAP5+ISW data combination (blue
dash-dot line). We can find that the combined constraint from
WMAP5+ISW is slightly improved over using WMAP5 alone, namely, the
$95\%$ upper limit of the coupling strength is $\beta<0.130$. Since
at present constraints from the ISW data are still very weak and in
the calculations we only consider the high-redshift quasar catalog
and totally neglect other low-redshift tracers which could give
powerful ISW constraints \cite{giannonho,hoetal}, the constraint on
$\beta$ does not improve significantly. When combining WMAP5, CMBExt
and ISW data together, the constraint on $\beta$ will improve
further: $\beta<0.120$ at $95\%$ confidence level.

Finally we combine all the current datasets together to constrain
the coupling strength $\beta$. In Fig.\ref{fig4} we find that the
constraint tightens significantly (magenta short dashed line):
\begin{equation}
\beta<0.050~(0.085)~\label{current}
\end{equation}
at $68\%$ ($95\%$) confidence level. The $95\%$ upper limit is
reduced by a factor of 2, when compared to the constraint from WMAP5
alone, which is due to the constraining power of SNIa and BAO on the
evolutions of cosmological background parameters. Meanwhile, the
parameter $\lambda$ in the potential form of quintessence scalar
field is constrained to be $\lambda<1.05$ at $95\%$ confidence
level, which is consistent with the result in Ref.\cite{bean08}.

\begin{figure}[t]
\begin{center}
\includegraphics[scale=0.37]{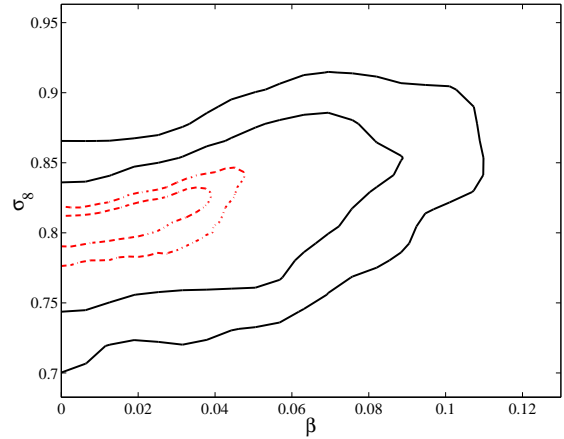}
\caption{Two dimensional marginalized contour in the ($\beta$,
$\sigma_8$) panel from all current data sets (black solid lines) and
all future data sets (red dash-dot lines),
respectively.\label{fig5}}
\end{center}
\end{figure}

As we mentioned before, the non-minimal coupling ($\beta>0$) affects
the large scale structure formation. From Fig.\ref{fig2} we can see
that the matter power spectrum on the small scales will be enhanced
significantly by the large value of $\beta$. The positive $\beta$
will lead to a high value of $\sigma_8$ today. Using the all
datasets combination, we obtain the limit on $\sigma_8$ today of
$\sigma_8=0.816\pm0.040$ ($68\%$ C.L.), which is obviously higher
than one obtained in the minimally coupled model ($\beta=0$):
$\sigma_8=0.793\pm0.030$ ($68\%$ C.L.), while the error bar is
enlarged slightly. Therefore, the $\sigma_8$ today and $\beta$ are
correlated as we expected, as shown the black solid lines in
Fig.\ref{fig5}.

\begin{figure}[t]
\begin{center}
\includegraphics[scale=0.37]{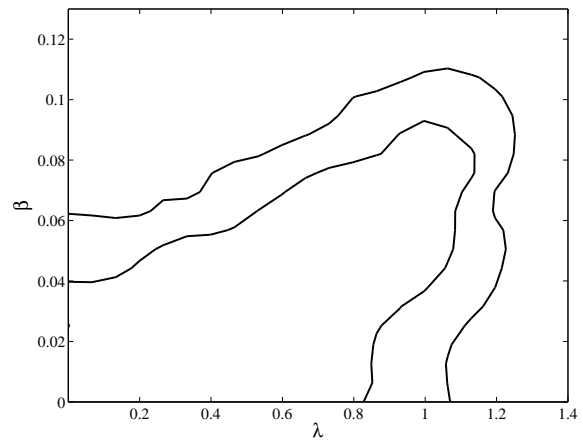}
\caption{Two dimensional marginalized contour in the ($\lambda$,
$\beta$) panel from all current data sets combination.\label{fig6}}
\end{center}
\end{figure}

In Fig.\ref{fig6} we also show the two dimensional contour in the
($\lambda$, $\beta$) panel. When $\lambda$ is larger than zero, the
equation of state of dark energy will evolve with cosmic time, and
the dark energy acts more like the dark matter ($w_{\phi}>-1$). In
this case the dark energy will dominate our Universe at relatively
higher redshift and slow down the linear growth function of the
matter perturbation and leads to a low value of $\sigma_8$ today.
The constraint of $\sigma_8=0.793\pm0.030$ ($68\%$ C.L.) from all
current datasets is obviously lower than that of the pure
$\Lambda$CDM: $\sigma_8=0.812\pm0.026$ ($68\%$ C.L.)
\cite{Komatsu:2008hk}. As the parameter $\lambda$ increases, in
order to reach the same value of $\sigma_8$ today, the strength of
coupling $\beta$ should be increased at the same time. Therefore,
$\lambda$ and $\beta$ are also correlated, illustrated in
Fig.\ref{fig6}.

From the results presented above, we can see the current
observations could give the constraints on the strength of
interaction as $\beta<\mathcal{O}(0.1)$ at $95\%$ confidence level.
It is worthwhile to discuss whether future data could give more
stringent constraints on $\beta$.

Therefore, in Fig.\ref{fig4} we also present the constraint on the
strength of interaction $\beta$ from the future datasets (navy blue
short dash-dot line). Due to the smaller error bars of the mock data
sets, the constraint on $\beta$ improves significantly, namely the
$68\%$ ($95\%$) upper limits are now
\begin{equation}
\beta<0.025~(0.039)~,
\end{equation}
which are reduced by another factor of 2, when compared to the
current constraint Eq.(\ref{current}). We also show the two
dimensional contour in the ($\beta$, $\sigma_8$) panel (red dash-dot
lines) in Fig.\ref{fig5}. The correlation still exists and the
limits have been shrunk significantly. These results imply that the
future measurements could constrain the non-minimal coupled dark
energy model with a higher precision.


\section{Summary}

The coupled dark energy models, in which the quintessence scalar
field non-minimally couples to the cold dark matter, could affect
the CMB temperature anisotropies and LSS matter power spectrum. In
this paper we present constraints on this coupled dark energy model
using the latest observations and future measurements. The WMAP5
data alone could only give a weak constraint on the strength of
coupling $\beta$. And then we exploit the capabilities of the
late-time ISW effect in constraining the non-minimal coupling, using
the cross-correlation signal between the quasar sample of SDSS DR6
\cite{richards09} and the WMAP5 ILC map \cite{hinshaw09}. We find
that the current ISW data could slightly improve the constraint on
$\beta$. If we add the BAO and SNIa data into our calculations, the
constraint on $\beta$ will improves significantly, namely the $95\%$
upper limit is $\beta<0.085$. Finally we simulate the future
measurements with smaller error bars and find that the future
measurements could improve the constraint on the strength of
coupling by a factor of two, when compared to the present
constraints.


\end{document}